\documentclass[cmp]{svjour}
\usepackage{graphics}
\usepackage{amsmath}
 \usepackage{epsfig}
\usepackage{amssymb}
\usepackage{amsfonts}
 \usepackage{latexsym}
\begin{document}

\def\pcs{P_c^\sharp}
\def\xa{\langle x \rangle}
\def\xam{\langle x\rangle^{-\sigma}} 
\def\xap{\langle x \rangle^{\sigma}}
\def\ta{\langle t \rangle}

\topmargin=2cm
\newtheorem{Proposition}{Proposition}
\newtheorem{Comment}{Proposition}
  \newtheorem{Remark}[Proposition]{Remark}
  \newtheorem{Corollary}[Proposition]{Corollary}
  \newtheorem{Lemma}[Proposition]{Lemma}
    \newtheorem{Theorem}[Proposition]{Theorem}
  \newtheorem{Note}[Proposition]{Note}
\newtheorem{Definition}{Definition}
\def\bfy{\mathbf{y}}
\def\bfz{\mathbf{z}}
\def\bfC{\mathbf{C}}
\def\Nsm{\hbox{\small I\hskip -2pt N}}
\def\rdb{\hbox{ I\hskip -2pt R}}
\def\cdb{\hbox{\it l\hskip -5.5pt C\/}}
\def\ndd{\hbox{\it I\hskip -2pt N}}
\def\zdd{\hbox{\sf Z\hskip -4pt Z}}
\def\e{\epsilon}
\def\ei{\epsilon^{-1}}
\def\a{\alpha}
\def\l{\lambda}
\def\tl{\tilde\lambda}
\def\ct{{\mbox{const}}}
\def\d{\delta}
\def\cf{{\cal F}}
    \def\z{\noindent}  
 \def\Box{{\hfill\hbox{\enspace${\sqre}$}} \smallskip}
    \def\sqr#1#2{{\vcenter{\vbox{\hrule height .#2pt
                             \hbox{\vrule width .#2pt height#1pt \kern#1pt
                                   \vrule width .#2pt}
                             \hrule height .#2pt}}}}
 \def\sqre{\mathchoice\sqr54\sqr54\sqr{4.1}3\sqr{3.5}3}     
     \def\erm{\mathrm{e}}
    \def\irm{\mathrm{i}}
    \def\drm{\mathrm{d}}
 \def\bchi{\mbox{\raisebox{.4ex}{\begin{Large}$\chi$\end{Large}}}}
     \def\CC{\mathbb{C}}
    \def\DD{\mathbb{D}}
    \def\NN{\mathbb{N}}
    \def\QQ{\mathbb{Q}}
    \def\RR{\mathbb{R}}
    \def\ZZ{\mathbb{Z}}

\title{Resonance Theory for Schr\"odinger Operators\\\ \\ \normalsize \it \centerline{Dedicated
to J. L. Lebowitz, on the occasion of his 70th birthday}}
\titlerunning{Resonance Theory}

\author{O. Costin\inst{1} \and
  A. Soffer\inst{1}\inst{1}
}                     
%
%
\institute{ Department of Mathematics\\ Rutgers University
\\ Piscataway, NJ 08854-8019}
\date{Received: date / Revised version: }
%
\maketitle

\begin{abstract}
  
Resonances which result from perturbation of embedded eigenvalues are
studied by time dependent methods.  A general theory is developed, with
new and weaker conditions, allowing for perturbations of threshold
eigenvalues and relaxed Fermi Golden rule.  The exponential decay rate
of resonances is addressed; its uniqueness in the time dependent picture
is shown is certain cases.  The relation to the existence of meromorphic
continuation of the properly weighted Green's function to time dependent
resonance is further elucidated, by giving an equivalent time dependent
asymptotic expansion of the solutions of the Schr\"odinger equation.
\end{abstract}

\section{Introduction and  results}
\label{intro}

\subsection{General remarks}Resonances may be defined in different ways, but usually refer to
metastable behavior (in time) of the corresponding system.  The standard
physics definition would be as ``bumps'' in the scattering cross
section, or exponentially decaying states in time, or poles of the
analytically continued $S$ matrix (when such an extension exists).

Mathematically, in the last 25 years one uses a definition close to
the above, by defining $\lambda$ to be a resonance (energy) if it is
the pole of the meromorphic continuation of the weighted Green's
function
$$
\chi(H-z)^{-1} \chi
$$
with suitable weights $\chi$ (usually, in the Schr\"odinger Theory
context, $\chi$ will be a $C_0^\infty$ function).  Here $H$ is the
Hamiltonian of the system.  In many cases the equivalence of some of the
above definitions has been shown \cite{G-Sig,H-S,Bal}.  However, the
exponential behavior in time, and the correct estimates on the remainder
are difficult to produce in general \cite{H-Sig}.  It is also not clear
how to relate the time behavior to a resonance, uniquely, and whether
``analytic continuation'' plays a fundamental role; see the review
\cite{Sim}.  Important progress on such relations has recently been
obtained; Orth \cite{Or} considered the time dependent behavior of
states which can be related to resonances without the assumption of
analytic continuation and established some preliminary estimates on the
remainder terms.  Then, Hunziker \cite{Hn} was able to develop a quite
general relation between resonances defined via poles of analytic
continuations in the context of Balslev-Combes theory, to exponential
decay in time, governed by the standard Fermi Golden rule.  Here the
resonances were small perturbations of embedded eigenvalues.  In
\cite{G-Sig} a definition of resonance in a time dependent way is given
and it is shown to agree with the one resulting from analytic
continuation when it exists, in the Balslev-Combes theory.  They also
get exponential decay and estimates on the remainder terms.

Exact solutions, including the case of large perturbations, for time
dependent potentials have recently been obtained in \cite{CLR}.  Further
notable results on the time dependent behavior of the wave equation were
proved by Tang and Zworski \cite{T-Z}. The construction of states which
resemble resonances, and thus decay approximately exponentially was
accomplished e.g. in \cite{Sk}.

For resonance theory based on Balslev-Combes method the reader is
referred to the book \cite{H-Sig} and its comprehensive bibliography on
the subject.

Then, in a time-dependent approach to perturbation of embedded
eigenvalues developed in \cite{SW} exponential decay and dispersive
estimates on the remainder terms were proved in a general context,
without the assumption of analytic continuation.  

When an embedded eigenvalue is slightly perturbed, we generally get a
``resonance''. One then expects the solution at time $t$ to be a sum of
an exponentially decaying term plus a small term (in the perturbation
size) which, however, decays slowly. 
The lifetime of the resonance is given by $\Gamma^{-1}$ where $\Gamma$,
the probability of decay per unit time, enters in the exponential decay
rate
$$p(t)\sim e^{-\Gamma t/\hbar}.$$
If an analytic continuation of $\chi(H_0-z)^{-1}\chi$ exists in a
neighborhood of an embedded eigenvalue, then $\Gamma=-2\Im z_0$, and a
resonance
$z_0$ is defined as the pole of the analytic continuation of 
$\chi(H-z)^{-1}\chi$. In this case, $\Gamma$ has the following expansion
in $\epsilon$

$$\Gamma(\lambda_0,\epsilon)=\epsilon^2\gamma(\lambda_0,\epsilon)+o(\epsilon^2)$$
The expression for $\gamma(\lambda_0,\epsilon)$ is called the Fermi
Golden Rule (FGR). A remarkable fact is that this expansion is defined
even when analytic continuation does not exist. Previous works on the
existence of resonances required that $\gamma(\lambda_0,\epsilon)>0$ as
$\epsilon\rightarrow 0$. This condition is sometimes hard to verify, and
in the present work we remove this assumption.

\subsection{Outline of new results} In this work we improve the theory of perturbation of embedded
eigenvalues and resonances in three main directions:

First, the Fermi Golden Rule condition, which originally required as
above (sometimes implicitly) that $\Gamma>C\epsilon^2$ as
$\epsilon\rightarrow 0$ is removed. We show that under (relatively weak)
conditions of regularity of the resolvent of the unperturbed Hamiltonian
all that is needed is that $\Gamma>0$. The price one sometimes has to
pay is that it may be needed to evaluate $\Gamma$ at a nearby point of
the eigenvalue $\lambda_0$ of the unperturbed Hamiltonian (see
(\ref{eq:(1)})). In cases of very low regularity of the unperturbed
resolvent, we need in general $\Gamma>C\epsilon^m$, with $m>2$; $m$
becomes larger if more regularity of the resolvent is provided; cf. 
(\ref{eq:g1}) and (\ref{eq:g2}) below.

The second main improvement relative to known results in resonance
theory is that we only require $H^\eta$ regularity (see \S\ref{def1}),
with $\eta>0$, of the unperturbed resolvent near the relevant
energy. Most works on resonance require analyticity; the recent works
\cite{Or,SW,H-Sig} require $H^\eta$ regularity with $\eta>1$. This
improvement is important to perturbations of embedded eigenvalues at
thresholds (e.g., our condition is satisfied by $H_0=-\Delta$ at
$\lambda_0=0$ in three or more dimensions, while the previous results
only apply to five or more dimensions).

As a third contribution we indicate that under conditions of analytic
continuation and with suitable cutoff, the term $e^{-\Gamma t}$ can be
separated from the solution and the remainder term is given by an
asymptotic series in $t^{-a}$, $a>0$, times a stretched exponential
$e^{-t^b}$, with $b<1$, see \S\ref{AC}.

Our analyticity assumptions are weaker and thus apply in cases of
threshold eigenvalues where standard complex deformation approaches
could fail. Furthermore we replace analytic perturbation methods by more
general complex theory arguments.

As concrete examples of applications we outline the following two
classes of problems;

\z (1) In many applications $H_0=-\Delta\oplus H_1$, where $H_1$ has a
discrete spectrum (see e.g. \cite{H-Sig}). if $H_1\psi_0=0$ has a
solution, then $H_0$ has an embedded eigenvalue at the threshold, since
$\sigma(-\Delta)=[0,\infty).$ In this case the known analytic methods do
not apply; the methods of \cite{Or} apply when $\eta>1$ which is the
case of the Laplacian on $L^2(\RR^N)$ if $N\ge 5$. The results of this paper
apply down to $N=3$.

\z (2) The Hamiltonians one gets by linearizing a nonlinear dispersive
completely integrable equation around an exact solution have an embedded
eigenvalue corresponding to the soliton/breather etc.  Small
perturbations of such completely integrable equations then produce a
perturbation problem of embedded eigenvalues with self-consistent
potential $W$. In these cases the size of $\Gamma$ is typically of
higher order in $\epsilon$ and in certain cases it is even
$O(e^{-1/\epsilon^2})$.  Hence the previous works are not applicable
since they require a lower bound $O(\epsilon^2)$ on $\Gamma$.

Our approach follows the setup of the time dependent theory of
\cite{SW}, combined with Laplace transform techniques.  It is expected
to generalize to the $N$-body case following \cite{M-S}.  We will
follow, in part, the notation of \cite{SW}. The analysis in this work
utilizes in some ways this framework, but generalizes the results
considerably: the required time decay is $O(t^{-1-\eta})$ and we remove
here the assumption of lower bound on $\Gamma$; it is replaced by

\begin{equation}
  \label{eq:g1}
 \Gamma \geq C\varepsilon^\frac{2}{1-\eta} 
\end{equation}

\z when $\eta < 1,$ and 
\begin{equation}
  \label{eq:g2}
\Gamma > 0, \text{ arbitrary}  
\end{equation}

\z when $\eta > 1$. 

  Whenever a meromorphic continuation of the $S$-matrix or Green's
function exists, the poles give an unambiguous definition of
``resonance.'' A time dependent approach or other definitions are less
precise, not necessarily unique, as was observed in \cite{Or}, but
usually apply in more general situations, where analytic continuation is
either hard to prove or not available.

We provide some information about defining resonance by time dependent
methods and its relation to the existence of ``analytic continuation''.

In particular, we will show that in general one can find the
exponential decay rate up to higher order corrections depending on
$\eta$ and $\Gamma$.

In case it is known that analytic continuation exists, our approach
provides a definition of a unique resonance corresponding to the perturbed
eigenvalue. It is given by the solution of some transcendental equation
in the complex plane and it also corresponds to a pole of the weighted
Green's function.

\section{Main results} We begin with some definitions.  Given $H_0$, a self-adjoint operator on 
$\mathcal{H}=L^2(\RR^n)$, we assume that $H_0$ has a simple eigenvalue $\lambda_0$
with normalized eigenvector $\psi_0$:

\begin{equation}
  \label{eq:(1)}
  H_0\psi_0=\lambda_0\psi_0, \|\psi_0\|=1
\end{equation}

\z Our interest is to describe the behavior of solutions of

\begin{equation}
  \label{eq:2}
  i\frac{\partial\phi}{\partial t}=H\phi, \ \ \ H:= H_0+\epsilon\, W^{(\epsilon)}
\end{equation}
where $\epsilon$ is a small parameter, taken to be the size of the
perturbation in an appropriate norm (cf. e.g. (\ref{eq:2.6})),
$\phi(0)=E_{\Delta}\phi_0$, where $E_\Delta$ is the spectral projection of
$H$ on the interval $\Delta$ and $\Delta$ is a small interval around
$\lambda_0$.  (Note that $W^{(\epsilon)}$ depends on $\epsilon$ in
general, and may not even have a limit as $\epsilon\rightarrow 0$.)
Furthermore, we will describe, in some cases, the analytic structure of
$(H-z)^{-1}$ in a neighborhood of $\lambda_0$. $W$ is a symmetric
perturbation of $H_0$, such that $H$ is self-adjoint with same domain as
$H_0$.

For an operator $A$, $\|A\|$ denotes its norm as an operator from $L^2$
to itself. We interpret functions of a self-adjoint operator as being 
defined by the spectral theorem. In the special case where the operator
is $H_0$, we omit the argument, i.e., $g(H_0)=g$.

For an open interval $\Delta$, we denote an appropriate smoothed
characteristic function of $\Delta$
by $g_\Delta(\lambda)$. In particular, we shall take typically  $g_\Delta(\lambda)$
to be a nonnegative
$C^\infty$ function, which is equal to one on $\Delta$ and zero outside
a neighborhood of $\Delta$. The support of its
derivative
is furthermore chosen to be  small compared to the size of $\Delta$.
We further require that $|g^{(n)}(\lambda)|\le c_n|\Delta|^{-n}, n\ge
1$.

$P_0$ denotes the projection on $\psi_0$, i.e., $P_0
f=(\psi_0,f)\psi_0$.
$P_{1b}$ denotes the spectral projection on $\mathcal{H}_{pp}\cap
\{\psi_0\}^\perp$,
the pure point spectral part of $H_0$ orthogonal to $\psi_0$. That is,
$P_{1b}$ projects onto the subspace of $\mathcal{H}$ spanned by all the
eigenstates other than $\psi_0$.
In our treatment, a central role is played by the subset of the spectrum
of the operator $H_0$, $T^\sharp$ on which a sufficiently rapid local
decay estimate holds. For a decay estimate to hold for $e^{-iH_0 t} $,
one must certainly project out the bound states of $H_0$, but there may
be other obstructions to rapid decay. In scattering theory these are
called threshold energies. Examples of thresholds are: (i)
points of stationary phase of a constant coefficient principal symbol
for two 
body Hamiltonians and (ii) for N--body Hamiltonians, zero and
eigenvalues of subsystems. We will not give a precise definition of
thresholds. For us it is sufficient to say that away from  thresholds
the favorable local decay estimates for $H_0$ hold.

Let $\Delta_*$ be a union of intervals, disjoint from $\Delta$,
containing a neighborhood of infinity and all thresholds of $H_0$ except
possibly those in a small neighborhood of $\lambda_0$. We then let

$$P_1=P_{1b}+g_{\Delta_*}$$

\z where $g_{\Delta_*}=g_{\Delta_*}(H_0)$ is a smoothed characteristic
function of the set 
$\Delta_*$. We also define for $x\in\RR^n$

\begin{equation}
  \label{eq:2.1}
  \langle x\rangle ^2=1+|x|^2,\ \ \ 
\overline{Q}=I-Q,\ \mbox{and}\ \ \ 
P_c^\sharp=I-P_0-P_1
\end{equation}

\z Thus, $\pcs$  is a smoothed out spectral projection of the set
$T^\sharp$
defined as 

\begin{equation}
  \label{eq:2.2}
  T^\sharp=\sigma(H_0)\setminus\{\mbox{eigenvalues, real neighborhoods
    of thresholds and infinity}\}
\end{equation}

\z We expect $e^{-iH_0 t} $ to satisfy good local decay estimates on the
range of $\pcs$; (see ({\bf H4}) below).

\bigskip

\subsection{Hypotheses on $H_0$}\label{def1} 
We assume $H^\eta$ regularity for $H_0$. By this  we
mean that $(\psi,(H_0-z)^{-1}\phi)$ is in the Sobolev space of order
$\eta$, $H^\eta$, in the $z$ variable for $z$ near the relevant
energy. Here $\psi,\phi$ are in the dense set
$\{\phi\in L^2:\langle x\rangle^{\sigma}\phi\in L^2\}$.

\bigskip

\z ({\bf H1}) $H_0$ is a self-adjoint operator with dense domain
$\mathcal{D}$, in $L^2(\RR^n)$.

\smallskip

\z ({\bf H2}) $\lambda_0$ is a simple embedded eigenvalue of $H_0$ with
(normalized) eigenfunction $\psi_0$.

\smallskip

\z ({\bf H3}) There is an open interval $\Delta$ containing 
$\lambda_0$ and no other eigenvalue of $H_0$.

\smallskip

\z ({\bf H4}) {\em Local decay estimate}: Let $r>1$. There exists
$\sigma>0$ such that if $\xa^\sigma f\in L^2$ then

\begin{equation}
  \label{eq:2.3}
  \| \xa^{-\sigma}e^{-iH_0t} \pcs f\|_2\le C\langle t\rangle ^{-r}\|
  \xa^\sigma f\|_2,
\end{equation}

\z ({\bf H5}) By appropriate choice of a real number $c$, the $L^2$
operator norm of 
$\xa^\sigma(H_0+c)^{-1}\xa^{-\sigma}$ can be made sufficiently small.

\smallskip

\bigskip

\z {\bf Remarks:}

\z (i) We have assumed that $\lambda_0$ is a simple eigenvalue to
simplify the presentation. Our methods
can be easily adapted to the case of multiple eigenvalues.

\smallskip

\z (ii) Note that $\Delta$ does not have to be small and that $\Delta_*$
can be chosen as necessary, depending on $H_0$. 

\smallskip

\z (iii) In certain cases, the above local decay conditions can be
proved even when $\lambda_0$ is a threshold; see \cite{JSS}. 

\smallskip

\z (iv) Regarding the verification of the local decay hypothesis, one
approach is to use techniques based on the Mourre estimate \cite{[26],[45],{Hun}}. If $\Delta$ contains no threshold values, then quite
generally,
the bound (\ref{eq:2.3}) holds with $r$ arbitrary and positive. 

\z We now specify the conditions we require of the perturbation, $W$.

\z {\bf Conditions on $W$}.

\z ({\bf W1}) $W$ is symmetric and $H=H_0+W$ is self-adjoint on
$\mathcal{D}$ and there exists $c\in\RR$ (which can be used in ({\bf
  H5})), such that $c$ lies in the resolvent sets of $H_0$ and $H$.

\z ({\bf W2}) For the same $\sigma$ as in ({\bf H4}) and ({\bf H5}) we
have :

\begin{multline}
  \label{eq:2.5}
  |||W|||:= \|\xa^{2\sigma}Wg_\Delta (H_0) \|\nonumber\\+
\|\xa^{\sigma}Wg_\Delta (H_0)\xa^{\sigma} \|+
\|\xa^{\sigma}W (H_0+c)^{-1}\xa^{-\sigma} \|<\infty
\end{multline}

\z and
\begin{equation}
  \label{eq:2.6}
\|\xa^{\sigma}W (H_0+c)^{-1}\xa^{\sigma} \|<\infty
\end{equation}

\z ({\bf W3}) {\em Resonance condition--nonvanishing of the Fermi golden
  rule}:

\z For a suitable choice of $\lambda$ (which will be made precise later)
\begin{equation}
  \label{eq:2.7} \Gamma(\lambda,\epsilon):= \Gamma(\lambda):= \pi\,
  \epsilon^2 (W^{(\epsilon)}\psi_0,\delta(H_0-\lambda)(I-P_0)W^{(\epsilon)}\psi_0)\ne 0
\end{equation}

\z In most cases $\Gamma=\Gamma(\lambda_0)$. But in the case 
$\Gamma$ is very small it turns out that the ``correct'' $\Gamma$ will be

$$\Gamma(\lambda_0+\delta)$$

\z with $\delta$ given in the proof of Proposition~\ref{roots}.
See also Section~\ref{S4}.

The main results of this paper are summarized in the following theorem.
\begin{Theorem}
  \label{2.1}
Let $H_0$ satisfy the conditions {\bf (H1)}...{\bf (H5)} and the perturbation
satisfy the conditions {\bf (W1)...(W3)}. Assume moreover that $\epsilon$ is
sufficiently small and either: 

\z (i) $H_0$ has regularity as in \S\ref{def1} with $\eta>1$ 

\smallskip

\z or

\smallskip

\z (ii) We have lower regularity $0<\eta<1$ supplemented by the conditions
$$\Gamma > C\epsilon^n,\ \ \ n\ge 2$$

\z and $\eta>\frac{n-2}{n}$.

\bigskip
\z 
Then

\z a) $H=H_0+\epsilon  W$ has no eigenvalues in $\Delta$.

\smallskip

\z b) The spectrum of $H$ is purely absolutely continuous in $\Delta$,
and 

\begin{equation}
  \label{eq:estn2}
  \|\xa^{-\sigma}e^{-iHt}g_{\Delta}(H)\Phi_0\|_2\le C_\epsilon \langle
  t\rangle ^{-1-\eta} \|\xa ^\sigma \Phi_0\|_2
\end{equation}

\smallskip

\z c) For $t\ge 0$ we have

\begin{equation}
  \label{eq:mainEst}
 e^{-iHt}g_\Delta(H)\Phi_0=(I+A_W)
\Big(e^{-i\omega_*t}a(0)\psi_0+ e^{-iH_0t}\phi_d(0)\Big)+R(t)
\end{equation}

\z where $A_W:=K(I-K)^{-1}-I$ and $K$ is an integral operator defined
in (\ref{eq:dK}) and

\begin{enumerate}
\item
if $\eta<1$ and $\epsilon\rightarrow 0$ with $t\Gamma$ fixed we have
$R(t)=O(\epsilon^2\Gamma^{\eta-1})$ while as $t\Gamma\rightarrow\infty$
we have $R(t)=O(\Gamma^{-1}t^{-\eta-1})$
\item
 for
$\eta>1$ we have
$R(t)=O(\epsilon^2 t^{-\eta+1})$ 
\item

 \begin{equation}\label{A}\|A_W\|\le C\epsilon |||W|||,\end{equation}
   $a(0)$ and $\phi_d(0)$
are determined by the initial data. The complex frequency $\omega_*$
is given by 

$$-i\omega_*=-is_0-\Gamma$$

\z where $s_0$ solves the equation

\begin{equation}
  \label{eq:w}
 s_0+\omega+\epsilon^2\Im\left\{F(\epsilon,is_0)\right\}=0
\end{equation}

\z (see (\ref{eq:laple}) and (\ref{eq:eql}) below) and 
\item
\begin{equation}
  \label{eq:defGamma}
  \Gamma=\epsilon^2\Re \left\{F(\epsilon,is_0)\right\}
\end{equation}

\end{enumerate}
\end{Theorem}

\z {\bf Remark:} $\omega_*$ can be found by solving the transcendental
equation (\ref{eq:w}) by either expansion or iteration if sufficient
regularity is present (see also Proposition~\ref{roots} and note
following it and Lemma~\ref{L3}).

\subsection{Sketch of the proof of the Theorem~\ref{2.1}}  The proof of
Theorem~\ref{2.1} is given in Secs.~\ref{S3} and \ref{S4}. Sec.~\ref{S3}
prepares the ground for the proof, Subsec.~\ref{4.1} provides key
definitions while Subsecs.~\ref{4.2} and \ref{Reg1} contain the proof of
Theorem~\ref{2.1} (ii) and (i) respectively.  As an intuitive guideline,
the solution $\phi(t)$ of the time dependent problem is decomposed into
the projection $a(t) \psi_0$ on the eigenfunction of $H_0$ and a
remainder (see (\ref{eq:3.1})). The remainder is estimated from the
detailed knowledge of $a(t)$ (see (\ref{eq:3.16}) and (\ref{eq:phid}).

Thus it is essential to control $a(t)$; once that is done, parts (a) and
(b) follow from the Proposition~\ref{PSW}; this $a(t)$ satisfies an
integral equation, cf. (\ref{eq:inta}). We chiefly use the Tauberian type
duality between the large $t$ behavior of $a(t)$ and the regularity
properties of its Laplace transform, cf. Proposition~\ref{Puniq} and
also Eq. (\ref{eq:ah}).
Then, an essential ingredient in the proof of the estimate
(\ref{eq:mainEst}) is Proposition~\ref{P0}. When enough regularity is
present, no lower bound on $\Gamma>0$ is imposed; Proposition~\ref{P15}
and Proposition~\ref{P16} are key ingredients here.

\subsection{Further results} 

\begin{Lemma} 
  \label{T2.3} Assuming the conditions of Theorem~\ref{2.1} with
$\eta>1$ then
\begin{equation} \label{eq:solw}
\omega_*=\lambda_0+\epsilon(\psi_0,W\psi_0)+(\Lambda+i\Gamma) +o(\epsilon^2)
\end{equation}

\z where
\begin{equation}
  \label{eq:Lambda}
  \Lambda=\epsilon^2(W\psi_0, P.V. (H_0-\lambda_0)^{-1}W\psi_0)
\end{equation}

\begin{equation}
  \label{eq:Gam}
  \Gamma=\pi\epsilon^2(W\psi_0,\delta(H_0-\lambda_0)(I-P_0)W\psi_0)
\end{equation}
\end{Lemma}

\z This follows from the proof of Proposition~\ref{Pr13} and the Remarks
below it.

\section{Decomposition and isolation of resonant terms}\label{S3}

We begin with the following decomposition of the solution of
(\ref{eq:2}):

\begin{eqnarray}
  \label{eq:3.1}
  e^{-iHt}\phi_0=\phi(t)=a(t)\psi_0+\tilde{\phi}(t)\\
\Big(\psi_0,\tilde{\phi}(t)\Big)=0,\ \ \ -\infty<t<\infty
\end{eqnarray}

\z Substitution into (\ref{eq:2}) yields

\begin{equation}
  \label{eq:3.3}
  i\partial_t\tilde{\phi}=H_0\phi+\e W\tilde{\phi}-(i\partial_t
  a-\lambda_0 a)\psi_0+a\e W\psi_0
\end{equation}

\z Recall now that $I=P_0+P_1+P_c^\sharp$. Taking the inner product of
(\ref{eq:3.3}) with $\psi_0$ gives the amplitude equation:

\begin{equation}
  \label{eq:3.4}
   i\partial_t a=(\lambda_0+(\psi_0,\e W\psi_0)\,)a+(\psi_0,\e
   WP_1\tilde{\phi})
+(\psi_0,\e W\phi_d),
\end{equation}

\z where

\begin{equation}
  \label{eq:3.5}
  \phi_d:= P_c^\sharp\tilde{\phi}
\end{equation}

\z The following equation for $\phi_d$ is obtained by applying
$P_c^\sharp$ to
equation (\ref{eq:3.3}):

\begin{equation}
  \label{eq:3.6}
   i\partial_t\phi_d=H_0\phi_d+P_c^\sharp\e W(P_1\tilde{\phi}+\phi_d)
+aP_c^\sharp\e W\psi_0
\end{equation}

\z To derive a closed system for $\phi_d(t)$ and $a(t)$ we now propose
to obtain an expression for $P_1\tilde{\phi}$, to be used in equations
(\ref{eq:3.4}) and (\ref{eq:3.6}). Since $g_\Delta(H)\phi(\cdot,t)=
\phi(\cdot,t)$ we find
\begin{equation}
  \label{eq:3.7}
  (I-g_\Delta(H))\phi=(I-g_\Delta(H))\Big( a\psi_0+P_1\tilde{\phi}+P_c^\sharp\tilde{\phi}\Big)=0
\end{equation}

\z or

\begin{equation}
  \label{eq:3.8}
 (I-g_\Delta(H)g_I(H_0)) P_1\tilde{\phi}=-\overline{g}_\Delta(H)\Big(a\psi_0+\phi_d\Big)
\end{equation}

\z where $g_I(\lambda) $ is a smooth function which is identically equal
to one on the support of $P_1(\lambda)$, and which has support disjoint
from $\Delta$. Therefore

\begin{equation}
  \label{eq:3.9}
  P_1\tilde{\phi}=-B\overline{g}_\Delta(H)(a\psi_0+\phi_d),
\end{equation}

\z where 

\begin{equation}
  \label{eq:3.9a}
  B=(I-g_\Delta(H)g_I(H_0))^{-1}.
\end{equation}

\z This computation is justified in Appendix B of \cite{SW}. The
following was also shown there:

\begin{Proposition}[\cite{SW}]
  \label{P3.1}

For small $\epsilon$, the operator $B$ in (\ref{eq:3.9a}) is a bounded
operator on $\mathcal{H}$.
\end{Proposition}

\bigskip

\z From (\ref{eq:3.9}) we get 

\begin{equation}
  \label{eq:3.10}
\phi(t)=a(t)\psi_0+\phi_d+P_1\tilde{\phi}=\tilde{g}_\Delta(H)(a(t)\psi_0+\phi_d(t)),
\end{equation}

\z with

\begin{equation}
  \label{eq:3.101}
  \tilde{g}_\Delta(H):= I-B\overline{g}_\Delta(H)=Bg_\Delta(H)(I-g_I(H_0)).
\end{equation}

\z see (\ref{eq:2.1}). Although $\tilde{g}_\Delta(H)$ is not really
defined as a function of $H$, we indulge in this mild abuse of notation
to emphasize its dependence on $H$. In fact, in
some sense, $\tilde{g}_\Delta(H)\sim g_\Delta(H) $ to higher order in
$\epsilon$ \cite{SW}.

Substitution of (\ref{eq:3.9}) into (\ref{eq:3.6}) gives:

\begin{equation}
  \label{eq:3.12}
  i\partial_t\phi_d=H_0\phi_d+aP_c^\sharp\e W\tilde{g}_\Delta(H)\psi_0+
P_c^\sharp\e W\tilde{g}_\Delta(H)\phi_d
\end{equation}

\z and

\begin{multline}
  \label{3.13}
i\partial_t a=\Big(\lambda_0+(\psi_0,\e W \tilde{g}_\Delta(H)\psi_0)
\Big)a+(\psi_0,\e W \tilde{g}_\Delta(H)\phi_d)\\=\omega a
+(\omega_1-\omega)a+
(\psi_0,\e W \tilde{g}_\Delta(H)\phi_d)
\end{multline}

\z where

\begin{eqnarray}
  \omega&=&\lambda_0+(\psi_0, \e W\psi_0)\\
\omega_1&=&\lambda_0+(\psi_0, \e W\tilde{g}_\Delta(H)\psi_0)
\end{eqnarray}

\z We write (\ref{eq:3.12}) as an equivalent integral equation. We will
later need the integral representation of the solution of (\ref{eq:3.12})

\begin{multline}
  \label{eq:3.16}
  \phi_d(t)=e^{-iH_0 t}\phi_d(0)-i\int_0^t e^{-iH_0(t-s)}a(s)P_c^\sharp\e
    W\tilde{g}_\Delta(H)\psi_0 ds\\-i\int_0^te^{-iH_0(t-s)}P_c^\sharp\e
    W \tilde{g}_\Delta(H)\phi_d ds
\end{multline}

\z This was also used to prove the following statement.
\begin{Proposition}[\cite{SW}]\label{PSW}
  Suppose $|a(t)|\le a_{\infty}\langle t\rangle^{-1-\alpha}$
and assume that $\eta>0$ and $\alpha\ge\eta$. Then for some $C>0$ we have 

$$\|\langle
x\rangle^{-\sigma}\phi_d(t)\|_{L^2}\le C \langle t\rangle^{-1-\eta}\left(
\|\langle
x\rangle^{\sigma}\phi_d(0)\|_{L^2}+a_{\infty}|||W|||\right)$$

\end{Proposition}

\z {\bf Note} The proposition, as we mentioned, implies parts (a) and
(b) of the main theorem, given the properties of $a(t)$ which will be
shown in the sequel. The absolute continuity stated in the theorem
follows from (\ref{eq:estn2}) with $\eta>0$.

We define $K$ as an operator acting on $C(\RR^+,\mathcal{H})$, the space
of continuous functions on $\RR^+$ with values in $\mathcal{H}$ by

\begin{equation}
  \label{eq:dK}
  \Big(K\, f\Big)(t,x)=\int_0^t e^{-iH_0(t-s)}P_c^\sharp\e
    W\tilde{g}_\Delta(H)f(s,x)ds
\end{equation}

\z We introduce on $C(\RR^+,\mathcal{H})$ the norm

\begin{equation}
  \label{eq:norm}
  \|f\|_\beta =\sup_{t\ge 0}\ta^\beta\|f(\cdot,t)\|_\mathcal{H}
\end{equation}

\z and define the operator norm

\begin{equation}
  \label{eq:normK}
  \|A\|_{\beta;\sigma}=\sup_{\|f\|_\beta \le 1}\|\xam A\xap f\|_\beta
\end{equation}

\z The above definitions directly imply the following.
\begin{Proposition}
  \label{eK}
If $\epsilon$ is small, $0\le \beta\le r$, $r>1$ and for some
$\beta_1>0$ we have
$\|\xa^{-\sigma}e^{-iH_0 t}P_c^\sharp \xa^{-\sigma} \|\le Ct^{-1-\beta_1}$, then for $0\le\beta\le\beta_1$ we have 

\begin{equation}
  \label{eq:nK}
  \|K f\|_{\beta;\sigma}\le \e\, C_{\beta;\sigma;r}
\end{equation}
\end{Proposition}

\z The proof uses the smallness of $\epsilon$ which in turn entails the
 boundedness of $\langle x\rangle^{-\sigma}\tilde{g}_\Delta(H)\langle 
x\rangle^{\sigma}$.  Using the definition of $K$ given above we see that
$K(1-K)^{-1}= \sum_{n=1}^{\infty}K^n$ is also bounded. We can now
rewrite the equations for $\phi_d$ as

\begin{multline}
  \label{eq:phid}
   \phi_d(t)=e^{-i H_0 t}\phi_d(0)+K\big(a(t)\psi_0\big)+K\phi_d\\=
(I-K)^{-1}\left\{K\big(a(t)\psi_0\big)+e^{-i H_0 t}\phi_d(0)\right\}
\end{multline}

\z (recall that we defined $A_W=-I+(I-K)^{-1}K$) and therefore

\begin{multline}
  \label{eq:eqa}
   i\partial_t a=\omega_1 a +\left(\psi_0,\e W
 \tilde{g}_\Delta(H)(I-K)^{-1}K\big(a\psi_0\big)\right)+\\
\left(\psi_0,\epsilon
 W\tilde{g}_\Delta(H)(I-K)^{-1}e^{-iH_0t}\phi_d(0)\right) 
\end{multline}

To complete the proof of Theorem~\ref{2.1} we need to estimate the large
time behavior of $a(t)$ solving Eq. (\ref{eq:eqa}). Since the
inhomogeneous term satisfies the required decay $O(t^{-1-\eta})$ by our
assumptions on $H_0$ it is sufficient to study the associated
homogeneous equation. Equivalently, we may choose the embedded
eigenfunction as initial condition (that is $\phi_d(0)=0$).

We now define two operators on $L^\infty$ by

\begin{equation}
  \label{eq:defjt}
  \tilde{j}(a)=\Big(v,\xam K(a\psi_0)\Big);\ \ \ \mbox{where}\ v=\xap \e
  W \tilde{g}_\Delta(H) \psi_0 
\end{equation}

\z and

\begin{equation}
  \label{eq:defj}
  {j}(a)=\Big(v,\xam (I-K)^{-1}K(a\psi_0)\Big)
\end{equation}

\begin{Proposition}
  \label{jj}
The operators $\tilde{j}$ and $j$ are bounded from $L^\infty$ into
itself.
\end{Proposition}

The proposition follows from Proposition~\ref{eK} with $\beta=0$.

\z {\bf Remark}. The equation for $a$ can now be written in the
equivalent integral
form

\begin{equation}
  \label{eq:inta}
  a(t)=a(0)e^{-i\omega t}+e^{-i\omega t}\int_0^t e^{i\omega s}j(a)(s)ds
:= a(0)e^{-i\omega t}+J(a)
\end{equation}
\begin{Definition}
  Consider the spaces $L^\infty_{T;\nu} $ and $L^\infty_{\nu} $
to be the spaces of functions on $[0,T]$ and $\RR^+$ respectively, in
the norm
\begin{equation}
  \label{eq:nnu}
  \|a\|_\nu =\sup_{s}|e^{-\nu s}a(s)|
\end{equation}

\begin{Remark}
  \label{Remn} 
  
  We note that for $T\in\RR^+$, the norm on $L^\infty_{T;\nu} $ is
  equivalent to the usual norm on $L^\infty[0,T]$.
\end{Remark}

\end{Definition}

\begin{Proposition}
 \label{Pnormj}
 For some constants ${c}$,   $C$ and $\tilde{c}$ {\em independent of
   $T$} we have $\|ja\|_\nu \le c\nu^{-1}\epsilon^2 \|a\|_\nu$, $\|J
 a\|_\nu \le C\nu^{-2}\epsilon^2 \|a\|_\nu$ and $\|\tilde{j}a\|_\nu \le
 \tilde{c}\nu^{-1}\epsilon^2\|a\|_\nu$, and thus $j$, $J$, and
 $\tilde{j}$ are defined on $L^\infty_{T;\nu}$ and $L^\infty_{\nu} $ and
 their norms, in these spaces, are estimated by

\begin{equation}
  \label{eq:njj}
  \|j\|_\nu\le c \nu^{-1}\epsilon^2 ;\ \ \|\tilde{j}\|_\nu \le
\tilde{c}\nu^{-1}\epsilon^2;\ \ \|J\|_\nu\le C\nu^{-2}\epsilon^2
\end{equation}

\end{Proposition}

\z Similar arguments as above lead to

\begin{Proposition}
 \label{Puniq}
 The equation (\ref{eq:eqa}) has a unique solution in
 $L^1_{loc}(\RR^+)$, and this solution belongs to $L^\infty_{\nu} $ if
 $\nu>\nu_0$ with $\nu_0$ sufficiently large. Thus, in the half-plane
 $\Re(p)>\nu_0$ the Laplace transform of $a$

\begin{equation}
  \label{eq:lapla}
  \hat{a}:= \int_0^{\infty} e^{-pt}a(t)dt
\end{equation}

\z exists and is analytic in $p$. Furthermore, for $\Re(p)>\nu_0$, the
Laplace transform of $a$ satisfies

\begin{equation}
  \label{eq:laple}
 ip\hat{a}=\omega\hat{a}+ia(0)-i\epsilon^2F(\epsilon,p)\hat{a}(p)
\end{equation}
where $F(\epsilon,p)$ is defined by 

\begin{multline}\label{eq:laple1}
F(\epsilon,p):=\\\left(\psi_0, W \tilde{g}_\Delta(H)\left[ \left(
      I+\frac{iI}{p+iH_0}P_c^\sharp
      W\tilde{g}_\Delta(H)\right)^{-1}\frac{-iI}{p+iH_0}P_c^\sharp\e W\tilde{g}_\Delta(H)
\right]\psi_0 \right)\\+i(\omega_1-\omega)\epsilon^{-2}
\end{multline}
so
\begin{equation}
  \label{eq:eql}
  (ip-\omega+i\epsilon^2F(\epsilon,p))\hat{a}(p)=ia(0)
\end{equation}
Eq. (\ref{eq:laple}) follows by taking the  Laplace transform of (\ref{3.13}).
\end{Proposition}

\begin{proof}
  By Proposition~\ref{Pnormj}, and since $\|e^{-i\omega t}\|_\nu=1$, for
  large $\nu$ the equation (\ref{eq:inta}) is contractive in
  $L^\infty_{T;\nu}$ and has a unique solution there. It thus has a
  unique solution in $L^1_{loc}$, by Remark~\ref{Remn}. Since by the
  same argument equation (\ref{eq:inta}) is contractive in
  $L^\infty_{T;\nu}$ and since $L^\infty_{\nu}\subset L^1_{loc}$, the
  unique $L^1_{loc}$ solution of (\ref{eq:inta}) is in $L^\infty_{\nu}$
  as well. The rest is straightforward.

\end{proof}

\begin{Remark}\label{DefF}
Note that by construction (\ref{eq:laple}) and (\ref{eq:laple1}) define
$F$ as a Laplace transform of a function.
\end{Remark}

Our assumptions easily imply that 
 if $\epsilon$ is small enough, then:

(a) $F(\epsilon,p)$ is analytic except for a cut 
along $i\Delta$. $F(\epsilon,p)$ is H\"older continuous of order $\eta>0$ 
 at the cut, i.e.

$$\lim_{\gamma\downarrow 0}F(\epsilon, i\tau\pm \gamma)\in H^\eta$$

\z the space of H\"older continuous functions of order $\eta$.

\smallskip

(b) $|F(\epsilon,p)|\le C|p|^{-1}$ for some $C>0$ as $|p|\rightarrow\infty$.

To
see it we write

  \begin{equation}
    \label{eq:B}
    B=B_1 B_2 \xam;\ \ B_1:= \frac{I}{p+iH_0}P_c^\sharp\xam ;\ \
    B_2:=
\e \xap W \tilde{g}_\Delta(H)\xap 
  \end{equation}

\z Noting that $P_c^\sharp$ projects on the interval $\Delta$ it is
clear
by the spectral theorem that $\xam B$ is analytic in $p$ on
$\mathcal{D}:=\CC\setminus(i\Delta)$. By the assumption on the decay rate
and the Laplace transform of eq. (\ref{eq:2.3}) we have that

\begin{equation}
  \label{eq:HC}
 B_3(p):= \xam\frac{I}{p+iH_0}P_c^\sharp \xam
\end{equation}

\z is uniformly H\"older continuous, of order $\eta$, as $p\rightarrow
i\Delta$. For $p_0\in i\Delta$, the two sided limits $\lim_{a\downarrow
  0}B_3(p_0\pm a)=B_3^\pm$ will of course differ, in general. A natural
closed domain of definition of $B_3$ is $\mathcal{D}$ together with the
two sides of the cut, $\overline{\mathcal{D}}:=
\mathcal{D}\cup\partial \mathcal{D}^+\cup \partial \mathcal{D}^-$. We
then write

\begin{equation}
  \label{eq:unifnorm}
  \left\| B_3 \right\|\le C_1(p)
\end{equation}

\z where we note that $C_1$ can be chosen so that:

\begin{Remark}
  \label{R3}
  $C_1(p)>0$ is uniformly bounded for $p\in\overline{\mathcal{D}}$ and
  $C_1(p)=O(p^{-1})$ for large $p$.
\end{Remark}

\z Hence for some $C_2$ we have
uniformly in $p$ (choosing $\e$ small enough),

\begin{equation}
  \label{eq:normB1B2}
  \|\xam (B_1 B_2)^n\|\le C_2^n\e^n
\end{equation}

\z and therefore the operator 

\begin{equation}
  \label{eq:op3}
  \e W \tilde{g}_\Delta(H)\left[ \left(
      I-\frac{I}{p+iH_0}P_c^\sharp\e
      W\tilde{g}_\Delta(H)\right)^{-1}\frac{I}{p+iH_0}P_c^\sharp\e W\tilde{g}_\Delta(H)
\right]
\end{equation}

\z is analytic in $\mathcal{D}$ and is
in $H^\eta(\overline{\mathcal{D}})$. 

\section{General Case}\label{S4}
\subsection{Definition of $\Gamma$}\label{4.1}

We have from Proposition~\ref{Puniq}, eq. (\ref {eq:laple}) that 

\begin{equation}
  \label{eq:ah}
  \hat{a}(p)=\frac{ia(0)}{ip-\omega+i\epsilon^2F(\epsilon,p)}
\end{equation}

\z We are most interested in the behavior of $\hat{a}$ for $p=is$,
$s\in\RR$. $\Gamma$ will be defined in terms of the approximate zeros of
the denominator in (\ref{eq:ah}).  Let $F=:F_1+iF_2$.

\begin{Proposition}\label{Pr13}
  \label{roots} For $\epsilon$ small enough, the equation
  $s+\omega+\epsilon^2 F_2(\epsilon,is)=0$ has at least one root $s_0$,
  and $s_0=-\omega+O(\epsilon^2)$. If $\eta\ge 1$, then for small enough
  $\epsilon$ the solution is unique. If $\eta<1$ then two solutions
  $s_1$ and $s_2$ differ by at most $O(\epsilon^{\frac{2}{1-\eta}})$.
\end{Proposition}

\begin{proof}
  We write $s=-\omega+\delta$ and get for $\delta$ an equation of the
  form $\delta=\epsilon^2G(\delta)$ where
  $G(x)=-F_2(\epsilon,ix-i\omega)$, and $G(x)\in H^\eta$. The existence
  of a solution for small $\epsilon$ is an immediate consequence of
  continuity and the fact that $\delta-\epsilon^2G(\delta)$ changes sign
  in an interval of size $\epsilon^2\|G\|_\infty$. If $\eta\ge 1$ we
  note that the equation $\delta=\epsilon^2G(\delta)$ is contractive for
  small $\epsilon$ and thus has a unique root. If instead $0<\eta<1$ we
  have, if $\delta_1,\delta_2$ are two roots, then for some $K>0$
  independent of $\epsilon$,
  $|\delta_1-\delta_2|=\epsilon^2|G(\delta_1)-G(\delta_2)|\le \epsilon^2
  K |\delta_1-\delta_2|^\eta$ whence the conclusion. $\Box$

\end{proof}

\begin{Remark}
  Note that $s_0$ are not, in general, poles of (\ref{eq:ah}) since we
  only solve for the real part equal to zero.
\end{Remark}
\z {\bf Assumption} If $\eta<1$ then we assume that $\epsilon^2
F_1(\epsilon,-i\omega)\gg \epsilon^{\frac{2}{1-\eta}}$ for small
$\epsilon$. When $\eta>1$ this restriction will not be needed,
cf. \S~\ref{Reg1}.

\medskip

\z {\bf Definition} We choose one solution
$s_0=-\omega+\delta$ and let $\Gamma$ be defined by (\ref{eq:defGamma}).

\z {\bf Note.}  
In the case $\eta<1$ the choice of $s_0$ yields, by the
previous assumption a (possible) arbitrariness in the definition of
$\Gamma$ of order $O(\epsilon^{\frac{2}{1-\eta}})=o(\Gamma)$.

\smallskip

\z {\bf Remarks on the verifiability of condition $\Gamma>0$}.  As it is
 generally difficult to check the positivity of $\Gamma$ itself but
 relatively easier to find $\Gamma_0$, we will look at various scenarios,
 which are motivated by concrete examples, in which the condition of
 positivity reduces to a condition on $F(\epsilon,-i\omega)$.

Let 

$$\Gamma_0=\epsilon^2 F_1(\epsilon, -i\omega);\ \ \ \gamma_0=\epsilon^2F_2(\epsilon,-i\omega)$$

\z where we see that $\Gamma_0$ and $\gamma_0$ are $O(\epsilon^2)$.
The equation for $\delta$ reads

$$\delta=-\epsilon^2
[F_2(\epsilon,-i\omega+i\delta)-F_2(\epsilon,-i\omega)]-\gamma_0=\epsilon^2
H(\delta)-\gamma_0$$

\z where $H(0)=0$. We write $\delta=-\gamma_0+\zeta$ and get

$$\zeta=\epsilon^2H(-\gamma_0+\zeta)$$

\z and the definition of $\Gamma$ becomes

$$\Gamma=\epsilon^2 F_1(\epsilon, -i\omega-i\gamma_0+i\zeta)$$

\begin{Proposition}\label{Pr15}
  (i) If $H_0$ satisfies the conditions of Theorem~\ref{2.1} with
$\eta>1$ and $\gamma_0=o(\epsilon^{-2}\Gamma_0)$, then
as $\epsilon\rightarrow 0$,
 
  \begin{equation}
    \label{eq:condG}
\Gamma=\Gamma_0+o(\Gamma_0)    
  \end{equation}

\z and in particular $\Gamma$ is positive for $\Gamma_0>0$.

(ii) Assume that $\eta<1$, $\gamma_0=o(\epsilon^{-2}\Gamma_0^{1/\eta})$ and
$\Gamma_0\gg
\epsilon^{\frac{2}{1-\eta}}$ as $\epsilon\rightarrow 0$. Then again
(\ref{eq:condG})
holds.

\end{Proposition}
\begin{proof}
(i) Since $\zeta=
O(\epsilon^2\gamma_0)+O(\epsilon^2\zeta)$ we get $\zeta=
O(\epsilon^2\gamma_0)$, implying that

$$\Gamma=\epsilon^2F_1\left[\epsilon, -i\omega-i\gamma_0(1+o(1))\right]
=\Gamma_0+O(\epsilon^2\gamma_0)=\Gamma_0+o(\Gamma_0)   $$

(ii) We have 

\begin{equation}
  \label{eq:111}
\zeta=O(\epsilon^2\gamma_0^\eta)+O(\epsilon^2\zeta^\eta)  
\end{equation}

If $\zeta\le \mbox{const}. \gamma_0$ as $\epsilon\rightarrow 0$, then the proof
is as in part (i). If on the contrary, for some large constant $C$ we
have $\zeta> C\gamma_0$ then by (\ref{eq:111}) we have $\zeta <
\mbox{const}.\epsilon^2\zeta^\eta$ so that $\zeta=O(\epsilon^{2/(1-\eta)})$ and
$\epsilon^2\zeta^\eta=O(\epsilon^{2/(1-\eta)})=o(\Gamma_0)$.  But then

$$\Gamma=\epsilon^2F_1(\epsilon, -i\omega)+O(\epsilon^2\gamma_0^\eta)+O(\epsilon^2\zeta^\eta)=\Gamma_0+o(\Gamma_0)$$

\end{proof}

\subsection{Exponential decay}~\label{4.2} We now let $p=is_0+v$. The intermediate
time and long time behavior of $a(t)$ are given by the following Proposition

\begin{Proposition}\label{P0}
For $t\Gamma=O(1)$ (note that $\Gamma$ in general depends on
$\epsilon$), as $\epsilon\rightarrow 0$ we have

(i)
\begin{equation}{\label{O}}
 a(t)=e^{-is_0 t}e^{-\Gamma
  t}+O(\epsilon^2\Gamma^{\eta-1})\end{equation}

(ii) As $t\rightarrow\infty$ we have 

\begin{equation}{\label{O*}} a(t)=O(\Gamma^{-1}t^{-\eta-1})\end{equation}
\end{Proposition}

\begin{proof}
  (i) Note first that, taking $\Re(v)>0$ and writing $F$ as a Laplace
  transform, cf. Remark~\ref{DefF}

$$F(\epsilon,-is_0+v)=\int_0^\infty
  e^{-is_0t-vt}f(t)dt$$

\z we have by our assumptions that

\begin{multline}
  \label{e}
  F(\epsilon,-is_0+v)=\int_0^\infty e^{-vt}\left(\int_0^t e^{-is_0
      u}f(u)du\right)'\\= v\int_0^\infty e^{-vt}\int_0^t e^{-is_0
    u}f(u)du = v\int_0^\infty
  e^{-vt}\left(\int_0^\infty-\int_t^\infty\right) e^{-is_0
    u}f(u)du\\=\int_0^\infty e^{-is_0 u}f(u)du-v \int_0^\infty
  e^{-vt}\int_t^\infty e^{-is_0 u}f(u)du\\=F(\epsilon,-is_0) -vL[g](v)
\end{multline}

\z where we denoted $g(v)=\int_t^\infty e^{-is_0 u}f(u)du$ and $L[g]$ is
its Laplace transform. Now define

\begin{equation}
  \label{eq:defh}
  h(v)=vL[g](v)
\end{equation}

\z We have, by the formula for the inverse Laplace transform 

\begin{equation}
  \label{eq:invlap}
  2\pi i a(t)=e^{-is_0 t}\int_{-i\infty}^{i\infty}\frac{e^{vt}}{v+\Gamma+\epsilon^2h(v)}dv
\end{equation}

\z where by construction we have $h\in H^\eta$, $h$ is analytic in
$\CC\setminus i\Delta$ and $h(0)=0$. We write

\begin{multline}
  \label{eq:expan}
 \int_{-i\infty}^{i\infty}\frac{e^{vt}}{v+\Gamma+\epsilon^2h(v)} dv
=\int_{-i\infty}^{i\infty}\frac{e^{vt}}{(v+\Gamma)\left(1+\epsilon^2
      h(v+\Gamma)^{-1}\right)}dv\\
=\int_{-i\infty}^{i\infty}\frac{e^{vt}dv}{v+\Gamma}-
\epsilon^2\int_{-i\infty}^{i\infty}\frac{1}{v+\Gamma}\frac{h(v+\Gamma)^{-1}}{1+\epsilon^2h(v+\Gamma)^{-1}}e^{vt}dv
\end{multline}

\z We first  need to estimate $L^{-1}\left[{h}(v+\Gamma)^{-1}\right]$
( the transformation is  well defined, since the function is just $(v+\Gamma)^{-1}(F(\epsilon,-is_0+v)-
F(\epsilon,-is_0)))$. We need to write 

\begin{equation}
  \label{eq:tr1}
vL[g](v)=:(v+\Gamma)L[g_1](v)  \ \ \mbox{or}\ \ L[g_1]=\left(1-\frac{\Gamma}{v+\Gamma}\right)L[g]
\end{equation}

\z which defines the function $g_1$:

\begin{equation}
  \label{a1}
  g_1=g-\Gamma e^{-\Gamma t}\int_0^t e^{\Gamma s}g(s)ds
\end{equation}

\z Since $|g(t)|<\mbox{Const.}t^{-\eta}$ we have

\begin{equation}
  \label{a4}
  |g_1(t)|\le \mbox{Const.}t^{-\eta}+e^{-\Gamma t}\int_0^{\Gamma
    t}e^u\left(\frac{u}{\Gamma}\right)^{-\eta}du\le 
\mbox{Const.} t^{-\eta}
\end{equation}

\z  A similar inequality holds for 

\begin{equation}
  \label{eq:Qdef}
  Q:=L^{-1}\left[\frac{\frac{
      h}{v+\Gamma}}{1+\frac{\epsilon^2
      h}{v+\Gamma}}\right]
\end{equation}

\z Indeed, we have

\begin{equation}
  \label{eq:Q}
 Q=-L^{-1}\left[\frac{h}{v+\Gamma}\right]+\epsilon^2L^{-1}\left[\frac{h}{v+\Gamma}\right]*Q
\end{equation}

\z It is easy to check that for $t\le r\Gamma^{-1}$ and small
enough $\epsilon$ this equation is
contractive in the norm $\|Q\|=\sup_{s\le t}\langle s\rangle^{\eta}
|Q(s)|$. 

But now, for constants independent of $\epsilon$, 
\begin{multline}
  \epsilon^2 L^{-1}\left[\frac{1}{v+\Gamma}\right]*Q\le \mbox{Const.} e^{-\Gamma
  s}\int_0^te^{\Gamma s}s^{-\eta} ds
\\=\epsilon^2\mbox{Const.} e^{-\Gamma
  s}\Gamma ^{-1}\int_0^{\Gamma t}e^{u}\left(\frac{u}{\Gamma}\right)^{-\eta} du
\le\mbox{Const.}\frac{\epsilon^2}{\Gamma^{1-\eta}}
\end{multline}

(ii)  We now use (\ref{e}) and (\ref{eq:defh}) to write

$$\frac{h}{v+\Gamma}=\frac{F(\epsilon,-is_0+v)}{v+\Gamma}-\frac{F(\epsilon,-is_0)}{v+\Gamma}$$

\z and get

$$ H_1:=L^{-1}\left[\frac{h}{v+\Gamma}\right]=e^{-\Gamma t}\int_0^t e^{\Gamma
  s}f(s)ds+conste^{-\Gamma t}$$

\z and thus, proceeding as in the proof of (i) we get for some $C>0$
$|H_1|\le C \Gamma^{-1}\langle t\rangle^{-\eta-1}$. To evaluate $a(t)$ for
large $t$ we resort again to $Q$ as defined in (\ref{eq:Qdef}) which
satisfies (\ref{eq:Q}). This time we note that the equation is
contractive in the norm $\sup_{s\ge 0}|\langle s\rangle^{1+\eta}\cdot |$
when $\epsilon$ is small enough. $\Box$
  
\end{proof}

 Using (\ref{O*}), Proposition~\ref{PSW} and (\ref{eq:3.10}) imply local
 decay and therefore $\chi$ cannot be an eigenfunction which implies
 (i). 
Since 
the local decay rate is integrable (ii) follows \cite{[RSIV]}. Part c)
 follows from (\ref{O}), (\ref{eq:phid}) and (\ref{eq:3.10})
 while (\ref{A}) follows from (\ref{eq:phid}) and the smallness of $K$.

\subsection{Proof of Theorem~\ref{2.1} in case (i) of regularity $\eta>1$}\label{Reg1} In this case we obtain  better
estimates.  We write

\begin{equation}
  \label{eq:defG}
G(v)=L^{-1}[g](v)  
\end{equation}

\z  and (\ref{eq:invlap}) becomes

\begin{equation}
  \label{f}
   a(t)=e^{-is_0 t}\int_{-i\infty}^{i\infty}\frac{e^{vt}}{v+\Gamma+\epsilon^2
     vG(v)}dv
\end{equation}

\z Now

\begin{multline}
  \label{g}
 L^{-1}\left[(v+\Gamma+\epsilon^2vG(v))^{-1}\right]\\
    =L^{-1}\left[\frac{1}{v+\Gamma}\right]-
\epsilon^2 L^{-1}\left[\frac{1}{v+\Gamma}\right]*
L^{-1}\left[\frac{\frac{v}{v+\Gamma}G(v)}{1+\epsilon^2\frac{v}{v+\Gamma}G(v)}\right]
\end{multline}

\begin{Proposition}\label{P15}
  Let 

$$H_2(t):=L^{-1}\left[\frac{\frac{v}{v+\Gamma}G(v)}{1+\epsilon^2\frac{v}{v+\Gamma}G(v)}\right]$$

\z We have 
\begin{equation}
  \label{h}
|H_2|\le \mbox{Const.} \langle 
t\rangle^{-\eta};\ \ \int_0^{\infty}H_2(t)dt=0  
\end{equation}

\end{Proposition}

\begin{proof}
  
 Consider first the function 

$$h_3: =v(v+\Gamma)^{-1}G(v)=G(v)-\Gamma(v+\Gamma)^{-1}G(v)$$

\z we see that (cf. (\ref{eq:defG}) and (\ref{e}))

\begin{multline}
H_3:=L^{-1}h_3=
\int_t^\infty e^{-is_0
  u}f(u)du-\Gamma\ e^{-\Gamma t}\int_0^t e^{\Gamma s}\int_s^\infty e^{-is_0
  u}f(u)du ds
\end{multline}

\z and thus, for some positive constants $C_i$,

\begin{equation}
  \label{eq:g}
|H_3|\le \mbox{Const.}t^{-\eta} +\mbox{Const.}e^{-\Gamma t}\int_0^{\Gamma
  t}e^v\Gamma^{-\eta}\langle v\rangle^{-\eta} dv\
\end{equation}

\z and thus, since $h_3(0)=0$ we have

$$|H_3|\le \mbox{Const.}\langle t\rangle^{-\eta};\ \ \int_0^{\infty}H_3(t)dt=0$$
\z Note now that the function

$$\frac{v}{v+\Gamma}G(v)\left(1+\epsilon^2\frac{v}{v+\Gamma}G(v)\right)^{-1}$$

\z vanishes for $v=0$. Note  furthermore that

$$H_2=H_3-\epsilon^2 H_3*H_2 $$

\z It is easy to check that this integral equation is contractive in the
norm $\|H\|=\sup_{s\le t}|\langle s\rangle^{\eta}H(s)|$ for small enough $\epsilon$;
the proof of the proposition is complete.
\end{proof}
\begin{Proposition}\label{P16}

$$L^{-1}\left[(v+\Gamma+\epsilon^2G(v))^{-1}\right]=e^{-\Gamma t}+\Delta(t)$$

\z where for some constant $C$ independent of $\epsilon, t, \Gamma$ we 
have

$$|\Delta|\le C\epsilon^2 \langle t\rangle^{-\eta+1}$$

\end{Proposition}

\begin{proof}
 
We have, by (\ref{g})

\begin{multline}
 \Delta(t) =\epsilon^2e^{-\Gamma t}\int_0^t e^{\Gamma
   s}\left(\int_s^{\infty}H_2(u)du\right)'ds\\=\epsilon^2\int_t^{\infty}H_2(s)ds-\Gamma e^{-\Gamma t}\int_0^t e^{\Gamma s}\int_s^{\infty}H_2(u)du
\end{multline}

\z The estimate of the last term is done as in (\ref{eq:g}). $\Box$
\end{proof}

\z Theorem~\ref{2.1} part (c) in case (i) follows.

\section{Analytic case}\label{AC}

Suppose that the function $F(p,\epsilon)$ has analytic continuation in a
neighborhood of the relevant 
energy $-i\omega\ne 0$; in this case we can prove stronger results. In many
cases one can show the analyticity of $F$ if the resolvent, properly
weighted, has analytic continuation.

\begin{Lemma}
  \label{L3} Assume that for some $\omega$ and some neighborhood
  $\mathcal{D}$ of $\omega$, $E(\epsilon,p)$ is a function with the following properties:

(i) $E\in H^\eta(\overline{\mathcal{D}}) $ and $E$ is analytic in
$\mathcal{D}$ (this allows for branch-points on the boundary of the
domain, a more general setting that meromorphicity).

(ii) $|E(\epsilon,p)|\le C\epsilon^2$ for some $C$.

(iii) $\lim_{a\downarrow 0} \Re E(\epsilon,-i\omega
-a)=-\Gamma_0<0$.

If (a) $\eta>1$, $E(\epsilon,-i\omega)=o(\Gamma_0/\epsilon^2)$ or (b)
$\eta<1$ and $E(\epsilon,-i\omega)=O(\Gamma_0)$ and $\epsilon$ is small
enough, then the function

$$G_1(\epsilon,p)=p+i\omega+E(\epsilon,p)$$

\z  has a
unique zero $p=p_z$ in $\overline{\mathcal{D}}$ and furthermore
$\Re(p_z)<0$. In fact,

\begin{equation}
  \label{eq:impz}
  \Re(p_z)+\Gamma_0=o(\Gamma_0)
\end{equation}

\end{Lemma}

\z {\bf Remark} If the condition that for $\eta>1$,
$E(\epsilon,-i\omega)=o(\epsilon^{-2}\Gamma_0)$ is not satisfied, then
we can replace $-i\omega$ by $-i\omega-is_0$ and the uniqueness of the
complex zero will still be true.

\begin{proof} We have

$$G_1(\epsilon,p_z)=0=p_z+i\omega+E(\epsilon,-i\omega)+[E(\epsilon,p_z)-E(\epsilon,-i\omega)]$$

\z or, letting $p=-i\omega+\zeta$, $\zeta_z:=p_z+i\omega$,
$\epsilon^2\phi(\epsilon,\zeta):=E(\epsilon,p)-E(\epsilon,-i\omega)$,

$$\zeta_z=-E(\epsilon,-i\omega)-\epsilon^2\phi(\epsilon,\zeta_z)$$

Consider a square centered at $ E(\epsilon,-i\omega)$ with side
$2|\Re(E(\epsilon,-i\omega))|=2\Gamma_0 $. For both cases (a) and (b)
for $\eta$ considered in part (iii) of the lemma, note that in our
assumptions and by the choice of the square we have

  \begin{equation} \label{eq:sqr}
    \left|\frac{\epsilon^2\phi(\zeta,\epsilon)}{\zeta+E(\epsilon,-i\omega)}\right|\rightarrow 0 \ \ (\mbox{as}\ \e\rightarrow 0)
    \end{equation} \z (on all sides of the square). In
    case (a) on the boundary of the rectangle we have by construction of
    the rectangle, $|\zeta+E(\epsilon,-i\omega)|\ge \Gamma_0$. Also by
    construction, on the sides of the rectangle we have
$|\zeta|\le \Gamma_0$. Still by assumption, $\phi(\epsilon,\zeta)\le
    C\zeta =o(\epsilon^{-2}\Gamma_0)$ and the ratio in (\ref{eq:sqr}) is
$o(1)$. In case (b), we have 

$$\epsilon^2\phi(\epsilon,\zeta) =O(\epsilon^2\zeta^\eta)=O(\epsilon^2\Gamma_0^\eta)=o(\Gamma_0)$$

Thus, on the boundary of the square, the variation of the argument of
    the functions $\zeta+E(\epsilon,-i\omega)+\e ^2\phi(\zeta)$ and that of
    $\zeta+E(\epsilon,-i\omega)$ differ by at most $o(1)$ and thus have to agree
    exactly (being integer multiples of $2\pi i$); thus
    $\zeta+E(\epsilon,-i\omega)+\e ^2\phi(\zeta)$ has exactly one root in the
    square.  The same argument shows that $p+i\omega+E(\epsilon,p)$ has
    no root in any other region in its analyticity domain except in the
    square constructed in the beginning of the proof. $\Box$

\smallskip

\end{proof}

\begin{Theorem}
  Assume the conditions (H) and (W) as before, and furthermore that the
  function $F(\epsilon,p)$ has analytic continuation in a neighborhood
  of $-i\omega$; with an appropriate choice of the cutoff function
  $E_\Delta(H_0)$, we have that $\chi (H-z)^{-1}\chi$ has a unique pole
  away from the real axis, near $-i\omega$, corresponding to a resonance
  with imaginary part near $\Gamma$, with appropriate choice of weights
  $\chi$.

\end{Theorem}
\begin{proof}First we note that by taking the Laplace transform of
(\ref{eq:3.10})
and (\ref{eq:3.16}) and solving for the resolvent of $H$ we get that 

$$\chi(H-z)^{-1}\chi =A(z)\hat{a}(z)\psi_0+B(z)$$

\z with $A(z)$ and $B(z)$ analytic in $\mathcal{D}$ by our assumptions
(H)
and (W), and the assumed analyticity of $F(\epsilon,p)$, $ip:=z$. 
Hence the existence and uniqueness of the pole of $\chi(H-z)^{-1}\chi$
follows from Lemma~\ref{L3}, with
$\epsilon^2F(\epsilon,p)=E(\epsilon,p)$.\end{proof}

As a consequence we obtain the following result.

\begin{Proposition}
  With an appropriate exponential cutoff function, the remainder
term decays as a stretched exponential times an asymptotic series.
\end{Proposition}

{\em Sketch of proof}. We need the large $t$ behavior of $a(t)$ which is
the Inverse Laplace transform of $G(p):=(p+i\omega+i\epsilon^2
F(\epsilon,p))^{-1}$ and to this end we write

\begin{equation}
  \label{eq:(11)}
  G(p)=(p+i\omega_*)^{-1}-i\epsilon^2(p+i\omega_*)^{-1}F_*(\epsilon,p)G(p)
\end{equation}

\z where $F_*(\epsilon,p):=F(\epsilon,p)-(\omega_*-\omega)/\epsilon^2$
and $\omega_*$ is the unique pole of $G(p)$ found in the previous
theorem.  Taking inverse Laplace transform of (\ref{eq:(11)}) we get an
integral equation for $G(t)$, and direct calculations show that
$\tilde{F}\sim e^{-\sqrt{t}+i\theta t}\sum a_k t^{-k/4}$ implies
$G(t)\sim e^{-i\omega_* t}+O(\epsilon^2) e^{-\sqrt{t}+i\theta t}\sum b_k
t^{-k/4}$. To find the asymptotic behavior of $\tilde{F}(t)$ we derive
an integral equation by taking the inverse Laplace transform of
(\ref{eq:laple1}) and the same integral equation arguments as above
reduce the asymptotic study of $\tilde{F}$ to that of the following 
expression for any $u\in L^2$:

$$(u,Be^{-iH_0 t} P_c^\sharp B\psi_0)=\int \widetilde{Bu^*}e^{-i\lambda
  t}g_{\Delta}\widetilde{B\psi_0}d\mu_{a.c.}(\lambda):= \int
\xi(\lambda)e^{-i\lambda t}g_{\Delta}(\lambda)d\lambda$$

\z where $B=W\tilde{g}_\Delta(H)$ and $\tilde{\phi}$ is the spectral
representation of $\phi$ associated to $H_0$. By assumption
$B(H_0-z)^{-1}B$ is analytic in $z\in\mathcal{D}$ hence
$\int(\widetilde{Bu^*})(\lambda)(\lambda-z)^{-1}(\widetilde{Bv})
(\lambda)f(\lambda)d\lambda$ is analytic for any $v\in L^2$, where
$f(\lambda)=d\mu_{a.c.}/d\lambda$; therefore so is its Hilbert transform
$\widetilde{Bu^*}\widetilde{Bv}f$ and thus $\xi$ is also
analytic. Choosing
$g_\Delta(\lambda)=\exp(-(\lambda-a)^{-1}+(\lambda-b)^{-1})$ the
asymptotic expansion of $\tilde{F}$ follows from that of the integral
$\int_a^b
e^{-\frac{1}{\lambda-a}+\frac{1}{\lambda-b}-it\lambda}\xi(\lambda)d\lambda$.
$\square$

\subsection{Example} Suppose

$$H_0=\begin{pmatrix}-\Delta&0\\0&-\Delta+x^2
\end{pmatrix}:=-\Delta\oplus (-\Delta+x^2)$$
on $L^2(\RR)\oplus L^2(\RR)$. Assume

$$W=\begin{pmatrix} 0&\tilde{W}\\\tilde{W} &0 \end{pmatrix}$$
with $\tilde{W}=\tilde{W}(x)$ sufficiently regular and exponentially
localized. Then, the spectrum
of $H_0$ has embedded eigenvalues corresponding to the spectrum of
$-\Delta+x^2$, with Gaussian localized and smooth eigenfunctions. 
Since the projection $I-P_0$ in the definition of $P_c^\sharp$
eliminates the $-\Delta+x^2$ part in any interval $\Delta$ containing 
an eigenvalue of $-\Delta+x^2$, it is left to verify the conditions of
the theorem
for $H_0$ replaced by $-\Delta$. 
Since

\begin{equation}\label{*1}
e^{-\alpha\langle x\rangle} (-\Delta-z)^{-1}e^{-\alpha\langle x\rangle}
\end{equation}
has analytic continuation through the cut $(0,\infty)$ and is an
analytic function
away from $z=0$, we can now choose an interval $\Delta=[a,b]$ around
each eigenvalue
$E_n$ of $-\Delta+x^2$, avoiding zero, and let

$$E_\Delta(\lambda)=e^{-(\lambda-a)^{-1}}e^{(\lambda-b)^{-1}}$$ a
function analytic in $\CC$ except $z=a$ and $b$.

\subsection{Remarks on applications} The examples covered by the above approach include those discussed in
\cite{SW} as well as the many cases where analytic continuation has been
established, see e.g. \cite{H-Sig}. Furthermore, following results of
\cite{H-Sig} it follows that under favorable assumptions on $V(x)$,
$-\Delta+V(x)$ has no zero energy bound states in three or more
dimensions extending the results of \cite{SW}, where it was proved for 5
or more dimensions.

 It is worth mentioning that the possible presence of thresholds inside
$\Delta$ makes it necessary to allow for $\eta<\infty$, and that in the
case where there are finitely many thresholds inside $\Delta$ of known
structure, sharper results may be obtained.

Other applications of our methods involve numerical reconstruction of
resonances from time dependent solutions data, in cases Borel
summability is ensured. This and other implications will be discussed
elsewhere.

\bigskip

\centerline{\bf \Large Acknowledgment}

\bigskip

The authors acknowledge partial support from the NSF. One of us (A. S.)
would like to thank I. M. Sigal for discussions.

\end{document}